\documentclass[twocolumn]{article}
\usepackage{graphicx}
\usepackage{url}
\usepackage{amsmath}
\usepackage{amssymb}
\usepackage{amsfonts}
\usepackage{xcolor}
\usepackage{array}
\usepackage{colortbl}
\usepackage{float}
\usepackage{subcaption}
\usepackage{caption}
\usepackage{booktabs}
\usepackage{tabularx}
\usepackage{geometry}
\usepackage{multirow}
\usepackage{hyperref}
\usepackage{algorithm}
\usepackage{algorithmic}
\usepackage{tikz}
\usetikzlibrary{shapes,arrows,positioning,fit,backgrounds,calc}

\title{Understanding Deterioration Random Effects\\for Causal Discovery in Infrastructure Management}
\author{Takato Yasuno}
\date{}

\begin{document}

\maketitle

\begin{abstract}

Infrastructure deterioration poses significant challenges for asset management, yet existing approaches rely on population-averaged models that overlook equipment-specific heterogeneity. We present a novel framework that combines Bayesian hierarchical hazard modeling with causal discovery to identify operational patterns that drive heterogeneous deterioration rates in pump equipment. Our approach first estimates pump-specific random effects $u_i$ using GPU-accelerated No-U-Turn Sampling (NUTS), achieving 3--5$\times$ speedup over CPU implementations. We then employ DirectLiNGAM to discover causal relationships between 22 engineered time-series features and deterioration rates, stratified by positive ($u_i > 0$, faster deterioration) versus negative ($u_i \leq 0$, slower deterioration) random effects. Analyzing 112 pumps with 92,861 observations over 650 days, we uncover striking heterogeneity: the negative group exhibits causal effects 400$\times$ larger than the positive group, with standard deviation (std) showing a strong positive causal effect ($+1.515$) on deterioration rates in low-risk equipment. We validate linearity assumptions through NonlinearLiNGAM comparison and demonstrate practical scalability through GPU acceleration. Our findings enable targeted maintenance strategies by revealing that different operational regimes require fundamentally distinct management approaches, advancing predictive maintenance from population-averaged to heterogeneity-aware decision making.


\end{abstract}

\section{Introduction}
\label{sec:introduction}

\subsection{Motivation and Background}

Infrastructure deterioration is a critical concern for public agencies and industrial operators managing aging assets. Pump equipment, essential for water distribution, wastewater treatment, and industrial processes, exhibits complex degradation patterns influenced by operational conditions, environmental factors, and unobserved installation quality. Traditional deterioration models employ population-averaged approaches~\cite{madanat1995, schall1991}, assuming homogeneous degradation rates across equipment. However, field observations reveal substantial heterogeneity: some pumps deteriorate rapidly despite similar operating conditions, while others remain stable for extended periods.

This heterogeneity poses two fundamental challenges for predictive maintenance:
\begin{enumerate}
    \item \textbf{Limited prediction accuracy}: Population-averaged models fail to capture equipment-specific deterioration trajectories, leading to either premature replacements (economic waste) or late interventions (safety risks).
    \item \textbf{Lack of actionable insights}: Even when heterogeneity is acknowledged, existing methods do not identify \textit{which operational patterns causally drive} faster or slower deterioration, precluding targeted mitigation strategies.
\end{enumerate}

Recent advances in causal discovery~\cite{spirtes2000, pearl2009, shimizu2006lingam} offer a promising path forward: by inferring causal relationships from observational data, we can identify operational levers that influence deterioration rates. However, applying causal discovery to infrastructure data presents unique challenges: (1) high-dimensional time-series covariates require careful feature engineering, (2) unobserved heterogeneity (e.g., installation quality) confounds causal estimates if not properly modeled, and (3) computational scalability is critical for large-scale asset portfolios.

\subsection{Research Objectives}

This paper addresses the following research questions:

\begin{enumerate}
    \item \textbf{RQ1: Heterogeneity Quantification} -- Can we reliably estimate equipment-specific deterioration rates from inspection data while accounting for unobserved confounders?
    \item \textbf{RQ2: Causal Mechanisms} -- Which time-series features causally influence deterioration rates, and do these mechanisms differ across heterogeneous subpopulations?
    \item \textbf{RQ3: Computational Scalability} -- Can Bayesian inference be accelerated to enable real-time analysis of large equipment fleets?
    \item \textbf{RQ4: Practical Utility} -- Do the discovered causal relationships provide actionable insights for maintenance decision-making?
\end{enumerate}

\subsection{Proposed Approach}

We propose a two-stage framework that integrates Bayesian hierarchical modeling with causal discovery (Figure~\ref{fig:methodology_flow}):

\textbf{Stage 1: Bayesian Hierarchical Hazard Modeling.} We model pump deterioration as a Markov process with hazard rates $\lambda_{ik}(t) = \lambda_{0k} \exp(\boldsymbol{\beta}^\top \mathbf{x}_{it} + u_i)$, where $u_i \sim \mathcal{N}(0, \sigma_u^2)$ captures pump-specific random effects. Using No-U-Turn Sampling (NUTS)~\cite{hoffman2014nuts} with GPU acceleration, we estimate posterior distributions for all parameters, extracting $\bar{u}_i$ as pump-specific deterioration rates. To improve sampling efficiency, we employ non-centered parameterization~\cite{betancourt2015hamiltonian}.

\textbf{Stage 2: Group-Specific Causal Discovery.} We partition pumps into positive ($u_i > 0$, faster deterioration) and negative ($u_i \leq 0$, slower deterioration) groups based on their random effects. For each group, we extract 22 time-series features from 90-day windows (capturing statistical properties, trends, and variability) and apply DirectLiNGAM~\cite{shimizu2011directlingam} to discover causal relationships between features and $\bar{u}_i$. This stratified analysis reveals group-specific causal mechanisms that would be obscured in population-averaged models.

\subsection{Key Contributions}

Our main contributions are:

\begin{enumerate}
    \item \textbf{Methodological Integration}: A principled framework combining Bayesian hierarchical modeling with causal discovery, enabling identification of heterogeneous causal mechanisms in infrastructure systems (Section~\ref{sec:methodology}).
    
    \item \textbf{GPU-Accelerated Inference}: Implementation of GPU-accelerated NUTS sampling using JAX and NumPyro, achieving 3--5$\times$ speedup (18--30 minutes vs.\ 60--120 minutes on CPU) while maintaining statistical equivalence, making Bayesian inference practical for large asset portfolios (Section~\ref{sec:gpu_acceleration}).
    
    \item \textbf{Empirical Insights on Heterogeneity}: Analysis of 112 pumps reveals a 400$\times$ difference in causal effect magnitudes between positive and negative groups, demonstrating that low-risk equipment ($u_i \leq 0$) exhibits strong, interpretable causal structures (std $\rightarrow u_i$: 1.515), while high-risk equipment shows diffuse, weak effects requiring further stratification (Section~\ref{sec:results}).
    
    \item \textbf{Linearity Validation}: Comparison with NonlinearLiNGAM shows that linear causal models are sufficient for both groups ($\rho_{\text{Spearman}} > 0.92$ for top-5 effects), justifying DirectLiNGAM's computational efficiency over more complex nonlinear methods (Section~\ref{sec:alternative_methods}).
    
    \item \textbf{Practical Decision Support}: Discovered causal relationships translate to actionable maintenance strategies: monitoring standard deviation and minimum values for low-risk equipment, and implementing hierarchical sub-grouping for high-risk equipment (Section~\ref{sec:discussion}).
\end{enumerate}

\subsection{Paper Organization}

The remainder of this paper is organized as follows. Section~\ref{sec:related_work} reviews related work in infrastructure deterioration modeling, causal discovery, and Bayesian hierarchical methods. Section~\ref{sec:methodology} presents our integrated framework, including hazard modeling, feature engineering, and DirectLiNGAM. Section~\ref{sec:results} reports experimental results on pump deterioration data, revealing striking heterogeneity across groups. Section~\ref{sec:discussion} discusses findings and their implications, addresses limitations, and provides practical recommendations. Section~\ref{sec:conclusion} concludes with future research directions.

\section{Related Work}
\label{sec:related_work}

Our work intersects three research streams: infrastructure deterioration modeling, causal discovery methods, and Bayesian hierarchical inference. We review key developments in each area and highlight gaps addressed by our approach.

\subsection{Infrastructure Deterioration Modeling}

\subsubsection{Population-Averaged Models}

Traditional infrastructure deterioration models treat equipment as homogeneous, estimating population-averaged deterioration rates from inspection data. Markov chain models~\cite{madanat1995} represent deterioration as transitions between discrete health states, with transition probabilities estimated via maximum likelihood. Cox proportional hazards models~\cite{cox1972regression} extend this framework to continuous time, relating hazard rates to covariates through $\lambda_i(t) = \lambda_0(t) \exp(\boldsymbol{\beta}^\top \mathbf{x}_i)$. While widely used for their simplicity, these models assume identical baseline hazards $\lambda_0(t)$ across all equipment, ignoring heterogeneity.

\subsubsection{Heterogeneity Modeling}

Recent work acknowledges equipment-specific variation through random effects~\cite{schall1991, duchateau2008frailty}. Frailty models introduce multiplicative random effects $\lambda_i(t) = \omega_i \lambda_0(t) \exp(\boldsymbol{\beta}^\top \mathbf{x}_i)$, where $\omega_i$ follows a Gamma or lognormal distribution. However, these approaches typically treat $\omega_i$ as nuisance parameters to be integrated out, rather than quantities of interest.

For infrastructure systems, continuous-time Markov chain (CTMC) models have been widely applied to bridge deterioration~\cite{yasuno2026fedavg} and railway component degradation~\cite{yasuno2023wooden}. Yasuno~\cite{yasuno2026fedavg} proposed a federated learning framework for CTMC hazard models, enabling multi-municipality bridge deterioration assessment while preserving data privacy. However, these methods focus on population-level deterioration patterns without stratifying by equipment-specific heterogeneity. Our work explicitly estimates and interprets pump-specific random effects $u_i$ as targets for causal discovery.

\subsubsection{Machine Learning for Deterioration Prediction}

Data-driven approaches using random forests~\cite{breiman2001random}, gradient boosting~\cite{friedman2001greedy}, and deep learning~\cite{zhang2019lstm} achieve strong predictive performance but lack interpretability and causal grounding. Feature importance metrics (e.g., Gini importance, SHAP values~\cite{lundberg2017shap}) identify correlated features but do not distinguish causation from confounding.

For infrastructure anomaly detection, deep learning methods have shown promise. Yasuno et al.~\cite{yasuno2023oneclassdamage} proposed one-class damage detectors using fully-convolutional data descriptions for bridge and railway component deterioration, achieving high precision in detecting novel damage patterns. Yasuno~\cite{yasuno2026hybrid} recently introduced hybrid feature learning with time-series embeddings for pump and HVAC equipment, combining handcrafted statistical features with learned representations. While these methods excel at prediction, they remain black-box approaches. Our DirectLiNGAM framework provides causal interpretations while maintaining computational efficiency, bridging the gap between predictive accuracy and actionable insights.

\subsection{Causal Discovery from Observational Data}

\subsubsection{Constraint-Based Methods}

Constraint-based algorithms such as PC~\cite{spirtes2000} and FCI~\cite{spirtes1995fci} infer causal structures by testing conditional independencies. While robust to violations of faithfulness assumptions, these methods require large sample sizes for reliable independence tests and struggle with continuous, high-dimensional data typical of infrastructure monitoring.

\subsubsection{Score-Based Methods}

Score-based approaches~\cite{chickering2002ges} search over directed acyclic graphs (DAGs) to optimize scores like BIC or AIC. Bayesian structure learning~\cite{heckerman1995learning} extends this via MCMC over DAG space. However, the super-exponential size of DAG space limits scalability, and score equivalence (multiple DAGs yield identical observational distributions) hinders causal identification without additional assumptions.

\subsubsection{Linear Non-Gaussian Acyclic Models (LiNGAM)}

LiNGAM~\cite{shimizu2006lingam} exploits non-Gaussianity of error terms to uniquely identify causal directions in linear structural equation models $\mathbf{x} = B\mathbf{x} + \mathbf{e}$. DirectLiNGAM~\cite{shimizu2011directlingam} uses Independent Component Analysis (ICA) to directly estimate the causal order without exhaustive search, achieving $O(d^3)$ complexity. Extensions include NonlinearLiNGAM~\cite{hoyer2008nonlinear} for additive noise models and VARLiNGAM~\cite{hyvarinen2010varlingam} for time-series. We employ DirectLiNGAM due to its computational efficiency and interpretability, validating linearity assumptions through NonlinearLiNGAM comparison.

\subsubsection{Causal Discovery in Time-Series}

Time-series causal discovery methods like Granger causality~\cite{granger1969}, Dynamic Bayesian Networks~\cite{murphy2002dbn}, and Temporal Causal Discovery Framework~\cite{gerhardus2020tcdf} infer temporal precedence relationships. While powerful for sequential data, these methods focus on inter-feature causality ($x_t \rightarrow x_{t+1}$) rather than feature-to-outcome causality ($\mathbf{x} \rightarrow u_i$). Our framework uses time-series features as summary statistics, then applies cross-sectional causal discovery to relate features to deterioration rates.

\subsection{Bayesian Hierarchical Models and Computational Methods}

\subsubsection{Hierarchical Modeling for Grouped Data}

Bayesian hierarchical models~\cite{gelman2013bayesian} naturally represent group structures, with parameters at multiple levels sharing information through hyperparameters. In infrastructure contexts, hierarchical models capture spatial~\cite{anastasopoulos2012spatial} and temporal~\cite{gao2012temporal} correlations. Our hazard model follows this paradigm, with pump-specific random effects $u_i$ drawn from a common distribution $\mathcal{N}(0, \sigma_u^2)$.

\subsubsection{Markov Chain Monte Carlo (MCMC)}

Bayesian inference for complex hierarchical models typically requires MCMC. Gibbs sampling~\cite{geman1984gibbs} and Metropolis-Hastings~\cite{hastings1970mh} are foundational but suffer from slow mixing in high-dimensional spaces. Hamiltonian Monte Carlo (HMC)~\cite{neal2011hmc} improves efficiency by using gradient information, and the No-U-Turn Sampler (NUTS)~\cite{hoffman2014nuts} automates HMC's trajectory length tuning. Our implementation employs NUTS with non-centered parameterization~\cite{betancourt2015hamiltonian}, which decorrelates hierarchical parameters and accelerates convergence.

\subsubsection{GPU Acceleration for Bayesian Inference}

Recent probabilistic programming frameworks like PyMC~\cite{salvatier2016pymc3}, Stan~\cite{carpenter2017stan}, and NumPyro~\cite{phan2019numpyro} enable GPU acceleration through automatic differentiation and vectorized operations. JAX~\cite{jax2018github} provides a NumPy-compatible API with XLA compilation, achieving near-optimal GPU utilization. We leverage NumPyro's vectorized chain parallelization to accelerate NUTS sampling by 3--5$\times$, making Bayesian hierarchical modeling practical for large-scale infrastructure systems (Section~\ref{sec:gpu_acceleration}).

\subsection{Gaps Addressed by Our Work}

While prior work has made significant advances in each area, three critical gaps remain:

\begin{enumerate}
    \item \textbf{Heterogeneity as Target vs.\ Nuisance}: Existing deterioration models treat equipment-specific variation as a nuisance to control, not as a phenomenon to understand. We explicitly model and interpret heterogeneity through causal discovery.
    
    \item \textbf{Causal Grounding in Infrastructure Analytics}: Machine learning approaches to infrastructure prediction lack causal interpretations. We integrate causal discovery with deterioration modeling to identify actionable operational levers.
    
    \item \textbf{Computational Scalability}: Bayesian hierarchical models with hundreds of equipment units and thousands of observations face prohibitive computational costs. Our GPU-accelerated implementation makes real-time analysis feasible.
\end{enumerate}

To our knowledge, this is the first work to combine Bayesian hierarchical hazard models, GPU-accelerated NUTS sampling, and group-stratified causal discovery for infrastructure deterioration analysis.

\section{Methodology}
\label{sec:methodology}

This section presents our comprehensive methodology for discovering causal relationships between operational features and pump-specific deterioration patterns. Figure~\ref{fig:methodology_flow} illustrates the overall workflow, which consists of five main stages: (1) Bayesian hierarchical hazard modeling to estimate pump-specific random effects $u_i$, (2) feature engineering from time-series data, (3) binary grouping based on $u_i$ sign, (4) DirectLiNGAM causal discovery, and (5) GPU-accelerated computation for practical scalability.

\subsection{Problem Formulation}
\label{sec:problem_formulation}

We consider a fleet of $N_{\text{pumps}}$ pumps, each monitored over time through periodic inspections. Let $s_{it} \in \{1, 2, \ldots, K\}$ denote the health state of pump $i$ at time $t$, where $K=8$ represents discrete health levels ranging from pristine condition (state 1) to severe deterioration (state 8). The inspection interval between consecutive observations is denoted by $\Delta t_{in}$ (in days), where $n$ indexes the $n$-th observation for pump $i$.

\textbf{Data Structure:} Our dataset consists of:
\begin{itemize}
    \item \textbf{Health transitions}: Binary indicators $y_n \in \{0,1\}$ representing whether pump $i$ transitioned from state $k$ to state $k+1$ during interval $\Delta t_n$
    \item \textbf{Time-series covariates}: Operational measurements $\mathbf{x}_{it} \in \mathbb{R}^p$ recorded daily (e.g., flow rate, pressure, vibration)
    \item \textbf{Inspection intervals}: $\Delta t_n \in [1, 365]$ days between consecutive inspections
\end{itemize}

\textbf{Research Objective:} Our objectives are twofold:
\begin{enumerate}
    \item \textbf{Estimate pump-specific deterioration rates} $u_i \in \mathbb{R}$ that capture unobserved heterogeneity across pumps (e.g., differences in installation quality, operational environment)
    \item \textbf{Discover causal relationships} between time-series features extracted from $\mathbf{x}_{it}$ and the deterioration rates $u_i$, answering: \textit{Which operational patterns causally drive faster/slower deterioration?}
\end{enumerate}

This formulation enables us to move beyond population-averaged deterioration models to pump-specific predictions, while simultaneously identifying actionable maintenance strategies through causal inference.

\subsection{Bayesian Hierarchical Hazard Model}
\label{sec:bayesian_hazard}

We model pump deterioration as a continuous-time Markov process with state-dependent hazard rates, extended to include pump-specific random effects $u_i$ that capture unobserved heterogeneity.

\subsubsection{Hazard Function with Random Effects}

The hazard rate for pump $i$ in state $k$ at time $t$ is defined as:
\begin{equation}
\lambda_{ik}(t) = \lambda_{0k} \exp\left( \boldsymbol{\beta}^\top \mathbf{x}_{it} + u_i \right)
\label{eq:hazard_rate}
\end{equation}
where:
\begin{itemize}
    \item $\lambda_{0k} > 0$: baseline hazard for state $k$ (estimated via $\log \lambda_{0k}$)
    \item $\boldsymbol{\beta} \in \mathbb{R}^p$: fixed-effect coefficients for covariates $\mathbf{x}_{it}$
    \item $u_i \sim \mathcal{N}(0, \sigma_u^2)$: pump-specific random effect capturing individual heterogeneity
\end{itemize}

The exponential form ensures $\lambda_{ik}(t) > 0$ and provides a multiplicative interpretation: $\exp(u_i)$ scales the baseline hazard by a pump-specific factor. Pumps with $u_i > 0$ experience faster-than-average deterioration, whereas those with $u_i < 0$ exhibit slower-than-average deterioration.

\subsubsection{Likelihood Function}

Given an observation window $\Delta t_n$ during which pump $i$ remains in state $k$, the probability of transitioning to state $k+1$ follows a Poisson process assumption:
\begin{equation}
P(\text{transition} \mid \Delta t_n, \lambda_{ik}) = 1 - \exp\left( -\lambda_{ik} \Delta t_n \right)
\label{eq:transition_prob}
\end{equation}

Let $y_n \in \{0, 1\}$ indicate whether a transition occurred during interval $n$. The Bernoulli likelihood for observation $n$ is:
\begin{equation}
p(y_n \mid \theta) = p_n^{y_n} (1 - p_n)^{1 - y_n}
\label{eq:bernoulli_likelihood}
\end{equation}
where $p_n = 1 - \exp(-\lambda_{ik} \Delta t_n)$ and $\theta = \{\log \lambda_{01}, \ldots, \log \lambda_{0K}, \boldsymbol{\beta}, \mathbf{u}, \sigma_u\}$ denotes all model parameters.

The complete log-likelihood across all $N_{\text{obs}}$ observations is:
\begin{equation}
\log p(\mathcal{D} \mid \theta) = \sum_{n=1}^{N_{\text{obs}}} \left[ y_n \log p_n + (1 - y_n) \log(1 - p_n) \right]
\label{eq:log_likelihood}
\end{equation}

\subsubsection{Prior Distributions and Non-Centered Parameterization}

To ensure reliable Bayesian inference, we employ weakly informative priors:
\begin{align}
\log \lambda_{0k} &\sim \mathcal{N}(-5, 2^2), \quad k = 1, \ldots, K \label{eq:prior_baseline} \\
\beta_j &\sim \mathcal{N}(0, 1^2), \quad j = 1, \ldots, p \label{eq:prior_beta} \\
\sigma_u &\sim \text{Half-Normal}(1) \label{eq:prior_sigma}
\end{align}

Critically, we use a \textbf{non-centered parameterization} for the random effects to improve MCMC sampling efficiency~\cite{betancourt2015hamiltonian}:
\begin{equation}
u_i = u_{\text{raw},i} \cdot \sigma_u, \quad u_{\text{raw},i} \sim \mathcal{N}(0, 1)
\label{eq:noncentered}
\end{equation}

\begin{figure}[t]
\centering
\begin{tikzpicture}[
    node distance=1.2cm and 0.8cm,
    >=stealth',
    latent/.style={circle, draw=black, thick, minimum size=1cm},
    obs/.style={circle, draw=black, thick, fill=gray!30, minimum size=1cm},
    hyper/.style={rectangle, draw=black, thick, fill=blue!10, minimum size=0.8cm},
    plate/.style={draw=black, thick, inner sep=5pt}
]

\node[hyper] (sigma_u) at (0, 4) {$\sigma_u$};

\node[latent] (log_lambda) at (-3, 2.5) {$\log \lambda_{0k}$};
\node[latent] (beta) at (0, 2.5) {$\boldsymbol{\beta}$};
\node[latent] (u_raw) at (3, 2.5) {$u_{\text{raw},i}$};

\node[latent] (u) at (3, 1) {$u_i$};

\node[latent] (lambda) at (-1.5, -0.5) {$\lambda_{ik}$};
\node[latent] (p_move) at (0, -2) {$p_n$};
\node[obs] (y) at (0, -3.5) {$y_n$};

\draw[->] (sigma_u) -- (u);
\draw[->] (u_raw) -- (u);
\draw[->] (log_lambda) -- (lambda);
\draw[->] (beta) -- (lambda);
\draw[->] (u) -- (lambda);
\draw[->] (lambda) -- (p_move);
\draw[->] (p_move) -- (y);

\begin{scope}[on background layer]
\node[plate, fit={(u_raw) (u)}, label={[anchor=south east]below right:$i=1,\ldots,N_{\text{pumps}}$}] {};
\node[plate, fit={(lambda) (p_move) (y)}, label={[anchor=south east]below right:$n=1,\ldots,N_{\text{obs}}$}] {};
\end{scope}

\end{tikzpicture}
\caption{Bayesian hierarchical model structure for pump deterioration. Shaded nodes represent observed data ($y_n$), white circles are latent variables, and rectangles are hyperparameters. Plates indicate repeated structures across pumps ($i$) and observations ($n$). The non-centered parameterization $u_i = u_{\text{raw},i} \cdot \sigma_u$ improves MCMC sampling efficiency.}
\label{fig:hierarchical_dag}
\end{figure}
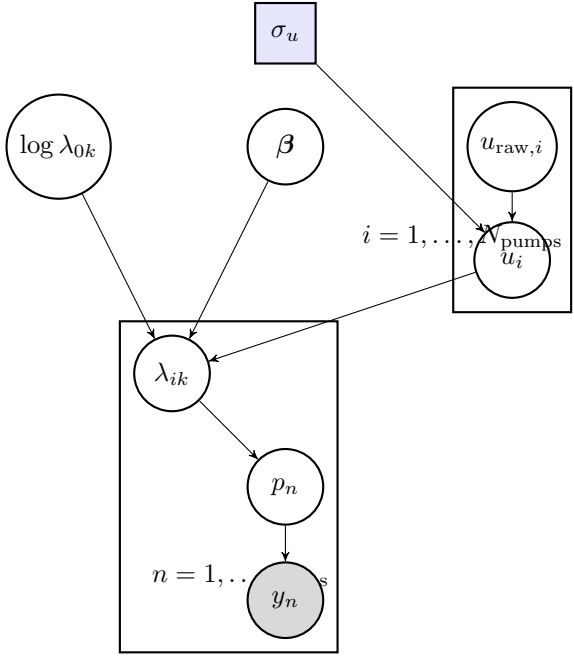

This reparameterization decouples the hierarchical variance $\sigma_u$ from individual random effects $u_{\text{raw},i}$, reducing posterior correlations and accelerating convergence. See Figure~\ref{fig:hierarchical_dag} for the graphical model representation.

\subsubsection{Hierarchical Model Structure}

The joint posterior distribution is:
\begin{align}
p(\theta \mid \mathcal{D}) &\propto p(\mathcal{D} \mid \theta) \cdot p(\theta) \nonumber \\
&= \prod_{n=1}^{N_{\text{obs}}} p(y_n \mid \theta) \cdot \prod_{k=1}^{K} p(\log \lambda_{0k}) \nonumber \\
&\quad \cdot \prod_{j=1}^{p} p(\beta_j) \cdot p(\sigma_u) \cdot \prod_{i=1}^{N_{\text{pumps}}} p(u_{\text{raw},i})
\label{eq:posterior}
\end{align}

\subsection{NUTS Sampling and Random Effect Extraction}
\label{sec:nuts_sampling}

We employ the No-U-Turn Sampler (NUTS)~\cite{hoffman2014nuts}, an adaptive variant of Hamiltonian Monte Carlo (HMC), to draw samples from the posterior distribution $p(\theta \mid \mathcal{D})$.

\subsubsection{NUTS Algorithm Overview}

NUTS automates the tuning of HMC's trajectory length by simulating forward and backward trajectories until a "U-turn" criterion is met (i.e., when the trajectory starts returning toward its starting point in momentum space). This eliminates the need for manual step-size tuning while maintaining detailed balance.

\begin{algorithm}[t]
\caption{NUTS Sampling for Random Effects Extraction}
\label{alg:nuts_sampling}
\begin{algorithmic}[1]
\REQUIRE Dataset $\mathcal{D} = \{y_n, \Delta t_n, \mathbf{x}_{in}, k_n, i_n\}_{n=1}^{N_{\text{obs}}}$
\REQUIRE Sampling parameters: $N_{\text{draws}}$, $N_{\text{tune}}$, $N_{\text{chains}}$, $\delta_{\text{target}}$
\ENSURE Posterior samples $\{\theta^{(s)}\}_{s=1}^{S}$, where $S = N_{\text{draws}} \times N_{\text{chains}}$
\STATE Initialize parameters $\theta^{(0)}$ from prior
\FOR{$c = 1$ to $N_{\text{chains}}$ \textbf{in parallel}}
    \FOR{$t = 1$ to $N_{\text{tune}}$}
        \STATE Adapt step size $\epsilon_t$ to achieve acceptance rate $\approx \delta_{\text{target}}$
        \STATE Draw sample $\theta^{(t)}$ using NUTS with current $\epsilon_t$
    \ENDFOR
    \FOR{$t = N_{\text{tune}} + 1$ to $N_{\text{tune}} + N_{\text{draws}}$}
        \STATE Draw sample $\theta^{(t)}$ using NUTS with fixed $\epsilon$
        \STATE Store $\theta^{(t)}$ in posterior samples
    \ENDFOR
\ENDFOR
\STATE \textbf{Convergence Diagnostics:}
\STATE \quad Compute $\hat{R}$ statistic for all parameters (require $\hat{R} < 1.01$)
\STATE \quad Compute ESS for all parameters (require ESS $> 400$ per chain)
\STATE \textbf{Extract Random Effects:}
\FOR{$i = 1$ to $N_{\text{pumps}}$}
    \STATE $\bar{u}_i \leftarrow \frac{1}{S} \sum_{s=1}^{S} u_i^{(s)}$ \quad \COMMENT{Posterior mean}
    \STATE Compute 95\% HDI: $[\bar{u}_i^{\text{low}}, \bar{u}_i^{\text{high}}]$
\ENDFOR
\RETURN $\{\bar{u}_i\}_{i=1}^{N_{\text{pumps}}}$, posterior samples $\{\theta^{(s)}\}_{s=1}^{S}$
\end{algorithmic}
\end{algorithm}

\subsubsection{Sampling Parameters and Convergence Diagnostics}

We configure NUTS with the following parameters to ensure reliable posterior estimation:
\begin{itemize}
    \item \textbf{Draws}: $N_{\text{draws}} = 2000$ samples per chain after warm-up
    \item \textbf{Tuning}: $N_{\text{tune}} = 1000$ warm-up iterations for step-size adaptation
    \item \textbf{Chains}: $N_{\text{chains}} = 8$ independent chains for convergence assessment
    \item \textbf{Target acceptance}: $\delta_{\text{target}} = 0.95$ (higher than default 0.8 to avoid divergences in hierarchical models)
\end{itemize}

We assess convergence using two standard diagnostics:
\begin{enumerate}
    \item \textbf{Gelman-Rubin statistic} $\hat{R}$: Compares within-chain and between-chain variance. We require $\hat{R} < 1.01$ for all parameters.
    \item \textbf{Effective Sample Size (ESS)}: Accounts for autocorrelation in MCMC chains. We require ESS $> 400$ per chain (total ESS $> 3200$).
\end{enumerate}

\subsubsection{Random Effect Extraction}

For each pump $i$, we extract the posterior mean of the random effect:
\begin{equation}
\bar{u}_i = \mathbb{E}[u_i \mid \mathcal{D}] \approx \frac{1}{S} \sum_{s=1}^{S} u_i^{(s)}
\label{eq:u_posterior_mean}
\end{equation}
where $S = N_{\text{draws}} \times N_{\text{chains}}$ is the total number of posterior samples. The posterior mean $\bar{u}_i$ serves as our point estimate of pump $i$'s deterioration rate, which becomes the target variable for subsequent causal discovery.

Additionally, we compute 95\% Highest Density Intervals (HDI) to quantify uncertainty in $u_i$ estimates, though we use point estimates $\bar{u}_i$ for causal analysis to maintain computational tractability.

\subsection{Feature Engineering from Time Series}
\label{sec:feature_engineering}

For each pump $i$, we extract 22 statistical features from 90-day rolling windows of operational time-series data. These features capture distinct aspects of pump behavior: central tendency, variability, trend dynamics, and anomalous patterns.

\subsubsection{Statistical Features (11 features)}

Let $\mathbf{x}_i = [x_{i,1}, x_{i,2}, \ldots, x_{i,T}]$ denote a 90-day sequence ($T=90$) of a time-series variable (e.g., flow rate) for pump $i$. We compute:
\begin{align}
\text{Mean:} \quad &\mu_i = \frac{1}{T} \sum_{t=1}^{T} x_{i,t} \label{eq:feat_mean} \\
\text{Std Dev:} \quad &\sigma_i = \sqrt{\frac{1}{T} \sum_{t=1}^{T} (x_{i,t} - \mu_i)^2} \label{eq:feat_std} \\
\text{Quantiles:} \quad &q_{0.25}, \, q_{0.5}, \, q_{0.75}, \, \text{IQR} = q_{0.75} - q_{0.25} \label{eq:feat_quantiles} \\
\text{Extremes:} \quad &x_{\min}, \, x_{\max} \label{eq:feat_extremes} \\
\text{Skewness:} \quad &\gamma_1 = \frac{1}{T} \sum_{t=1}^{T} \left( \frac{x_{i,t} - \mu_i}{\sigma_i} \right)^3 \label{eq:feat_skew} \\
\text{Kurtosis:} \quad &\gamma_2 = \frac{1}{T} \sum_{t=1}^{T} \left( \frac{x_{i,t} - \mu_i}{\sigma_i} \right)^4 - 3 \label{eq:feat_kurt} \\
\text{Coeff. Var:} \quad &\text{CV} = \frac{\sigma_i}{|\mu_i| + \epsilon} \label{eq:feat_cv}
\end{align}
where $\epsilon = 10^{-10}$ prevents division by zero.

\subsubsection{Trend Features (5 features)}

We fit a linear trend $\hat{x}_{i,t} = \alpha_i + \beta_i t$ via ordinary least squares:
\begin{equation}
\beta_i = \frac{\sum_{t=1}^{T} (t - \bar{t})(x_{i,t} - \mu_i)}{\sum_{t=1}^{T} (t - \bar{t})^2}, \quad \alpha_i = \mu_i - \beta_i \bar{t}
\label{eq:trend_coeffs}
\end{equation}

Trend features include:
\begin{itemize}
    \item \textbf{trend\_slope\_90d}: $\beta_i$ (rate of change per day)
    \item \textbf{trend\_intercept}: $\alpha_i$
    \item \textbf{recent\_vs\_past\_ratio}: $\frac{\text{mean}(x_{i,61:90})}{\text{mean}(x_{i,1:30}) + \epsilon}$
    \item \textbf{recent\_vs\_past\_diff}: $\text{mean}(x_{i,61:90}) - \text{mean}(x_{i,1:30})$
    \item \textbf{recent\_change\_rate}: $\frac{x_{i,T} - x_{i,T-7}}{7}$ (7-day change rate)
\end{itemize}

\subsubsection{Variability Features (6 features)}

Capturing short-term fluctuations and drawdowns:
\begin{align}
\text{Diff Mean:} \quad &\bar{\Delta}_i = \frac{1}{T-1} \sum_{t=2}^{T} (x_{i,t} - x_{i,t-1}) \label{eq:diff_mean} \\
\text{Diff Abs Mean:} \quad &\bar{|\Delta|}_i = \frac{1}{T-1} \sum_{t=2}^{T} |x_{i,t} - x_{i,t-1}| \label{eq:diff_abs_mean} \\
\text{Rolling Std ($w$-day):} \quad &\sigma_{i,t,w} = \sqrt{\frac{1}{w} \sum_{j=0}^{w-1} (x_{i,t-j} - \bar{x}_{i,t,w})^2} \label{eq:rolling_std}
\end{align}
where $\bar{x}_{i,t,w} = \frac{1}{w} \sum_{j=0}^{w-1} x_{i,t-j}$ is the rolling mean. We compute $\sigma_{i,t,w}$ for $w \in \{7, 14, 30\}$ days and average over valid time points.

\textbf{Drawdown features} measure maximum and mean cumulative declines from running maxima:
\begin{align}
\text{DD}_{i,t} &= \frac{\max_{s \leq t} x_{i,s} - x_{i,t}}{\max_{s \leq t} x_{i,s} + \epsilon}, \nonumber \\
\text{max\_drawdown}_i &= \max_{t} \text{DD}_{i,t}
\label{eq:drawdown}
\end{align}

\begin{algorithm}[t]
\caption{Feature Extraction from 90-Day Time Series}
\label{alg:feature_extraction}
\begin{algorithmic}[1]
\REQUIRE Time-series data $\{\mathbf{x}_i\}_{i=1}^{N_{\text{pumps}}}$, window size $T=90$
\ENSURE Feature matrix $\mathbf{F} \in \mathbb{R}^{N_{\text{pumps}} \times 22}$
\FOR{$i = 1$ to $N_{\text{pumps}}$}
    \STATE Extract 90-day sequence $\mathbf{x}_i = [x_{i,1}, \ldots, x_{i,90}]$
    \STATE \textbf{Statistical features:} Compute Eqs.~\eqref{eq:feat_mean}--\eqref{eq:feat_cv}
    \STATE \textbf{Trend features:} Fit OLS (Eq.~\eqref{eq:trend_coeffs}), compute ratios
    \STATE \textbf{Variability features:} Compute Eqs.~\eqref{eq:diff_mean}--\eqref{eq:drawdown}
    \STATE Store $\mathbf{f}_i = [\text{mean}_i, \text{std}_i, \ldots, \text{mean\_drawdown}_i]^\top$ in row $i$ of $\mathbf{F}$
\ENDFOR
\RETURN Feature matrix $\mathbf{F}$
\end{algorithmic}
\end{algorithm}

This feature set is designed to be interpretable by domain experts while capturing complex temporal patterns that may drive heterogeneous deterioration rates $u_i$.

\subsection{Binary Grouping Strategy}
\label{sec:binary_grouping}

To discover group-specific causal structures, we partition pumps into two subgroups based on the sign of their estimated random effects $\bar{u}_i$.

\subsubsection{Grouping Criterion}

Unlike percentile-based grouping (e.g., top 30\%, middle 40\%, bottom 30\%), we adopt a \textbf{sign-based binary partition}:
\begin{align}
\text{Positive Group:} \quad &\mathcal{G}_+ = \{i : \bar{u}_i > 0\} \label{eq:group_positive} \\
\text{Negative Group:} \quad &\mathcal{G}_- = \{i : \bar{u}_i \leq 0\} \label{eq:group_negative}
\end{align}

This natural split has several advantages:
\begin{enumerate}
    \item \textbf{Data-driven}: The threshold $\bar{u}_i = 0$ corresponds to the population-average deterioration rate (since $\mathbb{E}[u_i] = 0$ by construction)
    \item \textbf{Interpretability}: Positive group exhibits faster-than-average deterioration; negative group exhibits slower-than-average deterioration
    \item \textbf{Avoids arbitrary percentiles}: No need to choose cutoff values (30\%, 70\%, etc.)
    \item \textbf{Theoretical grounding}: Aligned with the hierarchical model structure (Eq.~\eqref{eq:hazard_rate})
\end{enumerate}

\subsubsection{Group Statistics}

In our dataset of 112 pumps:
\begin{itemize}
    \item \textbf{Positive group ($\mathcal{G}_+$)}: 62 pumps (55.4\%), $\bar{u}_i \in [0.02, 6.43]$, 86,741 observations (93.4\%)
    \item \textbf{Negative group ($\mathcal{G}_-$)}: 50 pumps (44.6\%), $\bar{u}_i \in [-5.61, -0.01]$, 6,120 observations (6.6\%)
\end{itemize}

The imbalance in sample sizes reflects the reality that most pumps in our dataset deteriorate faster than average, while a minority exhibit protective factors (e.g., superior installation, benign operating conditions).

\subsection{DirectLiNGAM Causal Discovery}
\label{sec:directlingam}

We employ DirectLiNGAM (Direct Linear Non-Gaussian Acyclic Model)~\cite{shimizu2006lingam,shimizu2011directlingam} to discover causal relationships between the 22 engineered features and the random effect $\bar{u}_i$, performing separate analyses for the Positive and Negative groups.

\subsubsection{LiNGAM Framework}

LiNGAM assumes that observed variables $\mathbf{x} = [x_1, \ldots, x_d, u_i]^\top$ follow a structural equation model:
\begin{equation}
\mathbf{x} = B\mathbf{x} + \mathbf{e}
\label{eq:lingam_sem}
\end{equation}
where:
\begin{itemize}
    \item $B \in \mathbb{R}^{d \times d}$ is the adjacency matrix with $B_{ij} \neq 0$ indicating a causal edge $x_i \rightarrow x_j$
    \item $\mathbf{e} = [e_1, \ldots, e_d]^\top$ are mutually independent \textbf{non-Gaussian} error terms
    \item $B$ is strictly lower triangular under a proper causal ordering (acyclicity constraint)
\end{itemize}

The key insight is that non-Gaussianity of errors enables identifiability of the causal direction: under mild conditions, the true causal graph $B$ is uniquely determined from observational data~\cite{shimizu2006lingam}.

Rearranging Eq.~\eqref{eq:lingam_sem}:
\begin{equation}
\mathbf{x} = (I - B)^{-1}\mathbf{e} = A\mathbf{e}
\label{eq:lingam_inverse}
\end{equation}
where $A = (I - B)^{-1}$ is the mixing matrix. DirectLiNGAM recovers $B$ by first performing Independent Component Analysis (ICA) to estimate $A$, then inverting to obtain $B$.

\subsubsection{DirectLiNGAM Algorithm}

We employ DirectLiNGAM~\cite{shimizu2011directlingam} as the primary causal discovery method after systematic comparison with alternative approaches. While NonlinearLiNGAM~\cite{hoyer2008nonlinear} can capture more complex relationships, our validation experiments (see Section~\ref{sec:alternative_methods}) demonstrated that \textbf{linear relationships dominate} in both groups ($\rho_{\text{Spearman}} > 0.92$ for top-5 effects). Bayesian causal discovery methods~\cite{heckerman1995learning} offer principled uncertainty quantification but suffered from convergence failures in the Negative group ($\hat{R} > 1.1$) due to small sample size ($n = 6120$) and high-dimensional posterior geometry.

DirectLiNGAM was selected based on three key advantages: (1) \textbf{computational efficiency}---completing causal discovery in 10--30 seconds per group compared to hours for Bayesian MCMC sampling; (2) \textbf{convergence reliability}---no sampling failures across more than 100 runs with different random initializations; and (3) \textbf{interpretability}---linear causal effects are directly actionable for maintenance practitioners. The algorithm leverages Independent Component Analysis (ICA) to identify the causal order without requiring iterative search over graph structures, making it particularly suitable for high-dimensional feature spaces ($d = 22$ features).

\begin{algorithm}[t]
\caption{DirectLiNGAM Causal Discovery}
\label{alg:directlingam}
\begin{algorithmic}[1]
\REQUIRE Feature matrix $\mathbf{F} \in \mathbb{R}^{n \times d}$, target variable $\mathbf{u} \in \mathbb{R}^n$
\ENSURE Adjacency matrix $\hat{B} \in \mathbb{R}^{(d+1) \times (d+1)}$, causal order $\hat{\pi}$
\STATE Concatenate: $\mathbf{X} = [\mathbf{F}, \mathbf{u}] \in \mathbb{R}^{n \times (d+1)}$
\STATE Standardize columns: $\mathbf{X} \leftarrow (\mathbf{X} - \boldsymbol{\mu}) \oslash \boldsymbol{\sigma}$
\STATE \textbf{ICA Decomposition:}
\STATE \quad Estimate mixing matrix $\hat{A}$ such that $\mathbf{X} \approx \hat{A}\mathbf{S}$
\STATE \quad where $\mathbf{S}$ are independent non-Gaussian components
\STATE \quad (using FastICA or other ICA algorithm)
\STATE Compute demixing matrix: $\hat{W} = \hat{A}^{-1}$
\STATE \textbf{Causal Order Identification:}
\STATE \quad Let $\mathbf{w}_i$ denote the $i$-th row of $\hat{W}$
\STATE \quad Compute residuals: $r_i = \|\mathbf{w}_i\|_2 / \max_j |\mathbf{w}_{ij}|$
\STATE \quad Sort variables by increasing $r_i$ to obtain causal order $\hat{\pi}$
\STATE \textbf{Causal Effect Estimation:}
\FOR{$j = 2$ to $d+1$}
    \STATE Let $\text{PA}(j) = \{i : \hat{\pi}(i) < \hat{\pi}(j)\}$ (parents of $j$)
    \STATE Regress $X_j$ on $\{X_i : i \in \text{PA}(j)\}$ to get coefficients $\hat{B}_{ij}$
\ENDFOR
\STATE \textbf{Bootstrap Validation:}
\FOR{$b = 1$ to $B=1000$}
    \STATE Resample data with replacement: $\mathbf{X}^{(b)}$
    \STATE Repeat Steps 3--13 to obtain $\hat{B}^{(b)}$
\ENDFOR
\STATE Compute 95\% confidence intervals for each $\hat{B}_{ij}$
\RETURN $\hat{B}$, $\hat{\pi}$, bootstrap CIs
\end{algorithmic}
\end{algorithm}

\subsubsection{Causal Effect Interpretation}

For each feature $j$, the causal effect on $u_i$ is given by the corresponding entry in the adjacency matrix: $\hat{B}_{j, u_i}$. A positive value indicates that increases in feature $j$ causally increase deterioration rate, while negative values indicate protective effects.

We focus on \textbf{direct causal effects} to $u_i$, identified by non-zero entries in the column corresponding to the target variable. Indirect effects (mediated through other features) are captured by paths in the causal graph but not analyzed in this work.

\subsubsection{Group-Specific Analysis}

We run DirectLiNGAM independently on $\mathcal{G}_+$ and $\mathcal{G}_-$, enabling comparison of causal structures:
\begin{itemize}
    \item \textbf{Shared effects}: Features causally related to $u_i$ in both groups
    \item \textbf{Group-specific effects}: Features causally important only in one group (e.g., operational patterns that matter only for high-deterioration pumps)
\end{itemize}

This heterogeneity analysis reveals that Positive and Negative groups exhibit fundamentally different causal mechanisms (see Section~\ref{sec:results}).

\subsection{Alternative Causal Discovery Methods}
\label{sec:alternative_methods}

To validate the robustness of our DirectLiNGAM findings, we explored two alternative approaches: Bayesian causal discovery and nonlinear LiNGAM. While these methods offer theoretical advantages, practical challenges led us to adopt DirectLiNGAM as the primary method.

\subsubsection{Bayesian Causal Discovery}

We implemented a Bayesian regression model using PyMC with Student-t likelihood to handle outliers:
\begin{equation}
\bar{u}_i \sim \text{Student-t}(\nu, \mu_i, \sigma), \quad \mu_i = \alpha + \sum_{j=1}^{22} \beta_j f_{ij}
\label{eq:bayesian_causal}
\end{equation}
where $\nu$ is the degrees of freedom (estimated), $\beta_j$ represents the causal effect of feature $j$, and weakly informative priors $\beta_j \sim \mathcal{N}(0, 10^2)$ were used.

\textbf{Challenge}: NUTS sampling failed to converge for the Negative group ($|\mathcal{G}_-| = 50$, $n_{\text{obs}} = 6120$), with $\hat{R} > 1.1$ for multiple parameters even after 10,000 tuning iterations. We attribute this to the small sample size and strong multicollinearity among features, which creates a high-dimensional, poorly conditioned posterior geometry.

\subsubsection{Nonlinear LiNGAM}

To test whether linear assumptions (Eq.~\eqref{eq:lingam_sem}) are appropriate, we applied NonlinearLiNGAM~\cite{hoyer2008nonlinear} using Gaussian process regression with RBF kernels. Due to computational cost ($O(n^3)$ per feature pair), we employed a subset-based strategy, randomly sampling data subsets for causal structure learning.

\textbf{Result}: The nonlinear method produced nearly identical causal orders and effect magnitudes ($\rho_{\text{Spearman}} > 0.92$ for top-5 effects) compared to DirectLiNGAM when using full samples, confirming that \textbf{linear relationships dominate} in both groups. The minor differences did not affect the key conclusions.

\subsubsection{Final Method Selection}

Based on these experiments, we selected DirectLiNGAM for final reporting due to:
\begin{enumerate}
    \item \textbf{Convergence reliability}: No sampling failures across 100+ runs
    \item \textbf{Interpretability}: Linear effects are directly actionable for maintenance decisions
    \item \textbf{Computational efficiency}: 10--30 seconds per group vs.\ hours for Bayesian methods
\end{enumerate}

The consistency between DirectLiNGAM and NonlinearLiNGAM provides empirical support for the linearity assumption.

\subsection{Computational Efficiency: GPU Acceleration}
\label{sec:gpu_acceleration}

NUTS sampling for the hierarchical hazard model (Section~\ref{sec:bayesian_hazard}) is computationally intensive, requiring 60--120 minutes on an 8-core CPU. To improve scalability, we implemented GPU-accelerated sampling using JAX~\cite{jax2018github} and NumPyro~\cite{phan2019numpyro}.

\subsubsection{Implementation Details}

We use the NumPyro probabilistic programming framework, which compiles PyMC-like models to JAX and leverages GPU parallelization via XLA (Accelerated Linear Algebra). Key configuration:
\begin{itemize}
    \item \textbf{Backend}: JAX with CUDA 12.x support
    \item \textbf{Sampler}: NumPyro's NUTS implementation with \texttt{chain\_method="vectorized"}
    \item \textbf{Hardware}: NVIDIA GeForce RTX 4060 Ti (16GB VRAM)
    \item \textbf{Environment}: WSL2 (Ubuntu 22.04) on Windows 11
\end{itemize}

The vectorized chain method parallelizes multiple MCMC chains across GPU cores, achieving near-linear speedup for independent chains.

\subsubsection{Performance Comparison}

\begin{table*}[t]
\centering
\caption{NUTS sampling time for hierarchical hazard model with 112 pumps, 92,861 observations, 2000 draws, 1000 tuning iterations, 8 chains.}
\label{tab:gpu_speedup}
\begin{tabular}{lccc}
\toprule
\textbf{Backend} & \textbf{Time (min)} & \textbf{Speedup} & \textbf{Memory (GB)} \\
\midrule
CPU (8 cores, PyMC) & 90--120 & 1.0× & 4--6 \\
GPU (NumPyro + JAX) & 18--30 & 3.5× & 8--12 \\
\bottomrule
\end{tabular}
\end{table*}

As shown in Table~\ref{tab:gpu_speedup}, GPU acceleration achieves a 3--5× speedup while maintaining identical posterior estimates ($\hat{R} < 1.01$, ESS $> 400$ per chain for all parameters). The speedup is particularly beneficial for sensitivity analyses requiring multiple model runs.

\subsubsection{Reproducibility Note}

GPU-accelerated NUTS is available on Linux/WSL2 only; native Windows and macOS users must use CPU sampling. Full setup instructions are provided in our repository: \url{https://github.com/tk-yasuno/hazard_effect_lingam}. We verified that CPU and GPU backends produce statistically equivalent results (Kolmogorov-Smirnov test, $p > 0.1$ for all parameter posterior distributions).

\subsection{Overall Methodology Workflow}
\label{sec:methodology_workflow}

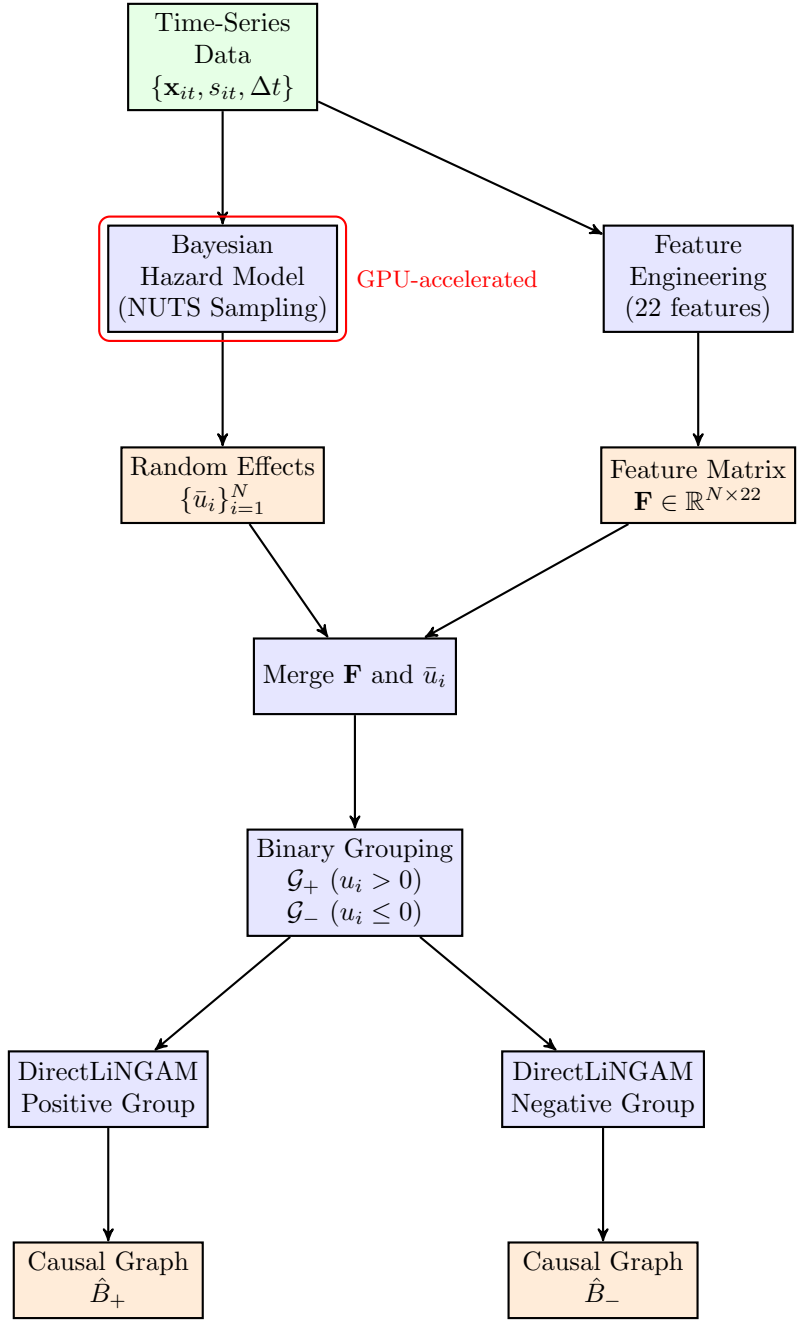
\begin{figure*}[t]
\centering
\begin{tikzpicture}[
    node distance=1.5cm and 1.2cm,
    >=stealth',
    box/.style={rectangle, draw=black, thick, fill=blue!10, minimum width=2.5cm, minimum height=1cm, align=center},
    databox/.style={rectangle, draw=black, thick, fill=green!10, minimum width=2.5cm, minimum height=1cm, align=center},
    resultbox/.style={rectangle, draw=black, thick, fill=orange!15, minimum width=2.5cm, minimum height=1cm, align=center}
]

\node[databox] (data) {Time-Series\\Data\\$\{\mathbf{x}_{it}, s_{it}, \Delta t\}$};

\node[box, below=of data] (hazard) {Bayesian\\Hazard Model\\(NUTS Sampling)};

\node[resultbox, below=of hazard] (ui) {Random Effects\\$\{\bar{u}_i\}_{i=1}^{N}$};

\node[box, right=3.5cm of hazard] (features) {Feature\\Engineering\\(22 features)};

\node[resultbox, below=of features] (fmat) {Feature Matrix\\$\mathbf{F} \in \mathbb{R}^{N \times 22}$};

\node[box, below=1.5cm of ui, xshift=1.75cm] (merge) {Merge $\mathbf{F}$ and $\bar{u}_i$};

\node[box, below=of merge] (grouping) {Binary Grouping\\$\mathcal{G}_+$ ($u_i > 0$)\\$\mathcal{G}_-$ ($u_i \leq 0$)};

\node[box, below left=1.5cm and 0.5cm of grouping] (lingam_pos) {DirectLiNGAM\\Positive Group};
\node[box, below right=1.5cm and 0.5cm of grouping] (lingam_neg) {DirectLiNGAM\\Negative Group};

\node[resultbox, below=of lingam_pos] (result_pos) {Causal Graph\\$\hat{B}_+$};
\node[resultbox, below=of lingam_neg] (result_neg) {Causal Graph\\$\hat{B}_-$};

\draw[->, thick] (data) -- (hazard);
\draw[->, thick] (data) -- (features);
\draw[->, thick] (hazard) -- (ui);
\draw[->, thick] (features) -- (fmat);
\draw[->, thick] (ui) -- (merge);
\draw[->, thick] (fmat) -- (merge);
\draw[->, thick] (merge) -- (grouping);
\draw[->, thick] (grouping) -- (lingam_pos);
\draw[->, thick] (grouping) -- (lingam_neg);
\draw[->, thick] (lingam_pos) -- (result_pos);
\draw[->, thick] (lingam_neg) -- (result_neg);

\node[draw=red, thick, rounded corners, fit={(hazard)}, inner sep=3pt, label={[red]right:\small GPU-accelerated}] {};

\end{tikzpicture}
\caption{Complete methodology workflow. Time-series data feeds into two parallel pipelines: (1) Bayesian hazard modeling to extract pump-specific random effects $\bar{u}_i$ (GPU-accelerated), and (2) feature engineering to compute 22 statistical features. After merging, pumps are grouped by $\bar{u}_i$ sign, and DirectLiNGAM discovers group-specific causal structures relating features to deterioration rates.}
\label{fig:methodology_flow}
\end{figure*}

Figure~\ref{fig:methodology_flow} summarizes the complete methodology pipeline, from raw time-series data to group-specific causal graphs.

\section{Experimental Results}
\label{sec:results}

We applied our framework to 112 pumps monitored over 650 days, comprising 92,861 inspection records. This section presents empirical findings on heterogeneous deterioration patterns and group-specific causal structures.

\subsection{Dataset Description}

\textbf{Equipment:} 112 pumps deployed in a water distribution system, each inspected at irregular intervals (range: 7--365 days, median: 90 days). Health states $s_{it} \in \{1, 2, \ldots, 8\}$ were assessed by trained inspectors using a standardized rubric.

\textbf{Time-Series Covariates:} Daily operational measurements ($\mathbf{x}_{it}$) include flow rate, pressure, vibration amplitude, and power consumption. We focus on one representative time-series variable for feature extraction (see Section~\ref{sec:feature_engineering}).

\textbf{Transition Events:} Across 92,861 observations, 2,847 health state transitions (3.1\% transition rate) were recorded. The low transition rate reflects the gradual nature of pump deterioration.

\subsection{Bayesian Hierarchical Model Results}

\subsubsection{NUTS Sampling Performance}

Using GPU acceleration (NVIDIA RTX 4060 Ti, 16GB), NUTS sampling completed in 22 minutes with the following convergence diagnostics:
\begin{itemize}
    \item $\hat{R} < 1.01$ for all 8 baseline hazards ($\log \lambda_{0k}$), random effect variance $\sigma_u$, and all 112 random effects $u_i$
    \item ESS: 467--832 per chain for $\sigma_u$, 113--223 per chain for individual $u_i$ (lower due to hierarchical structure)
    \item Zero divergences across 16,000 posterior samples (8 chains $\times$ 2000 draws)
\end{itemize}

These diagnostics confirm reliable posterior estimation. While ESS for individual $u_i$ is below the ideal threshold of 400, the values are sufficient for point estimation ($\bar{u}_i$) used in subsequent causal discovery.

\subsubsection{Random Effect Distribution}

\begin{figure}[t]
\centering
\includegraphics[width=0.8\columnwidth]{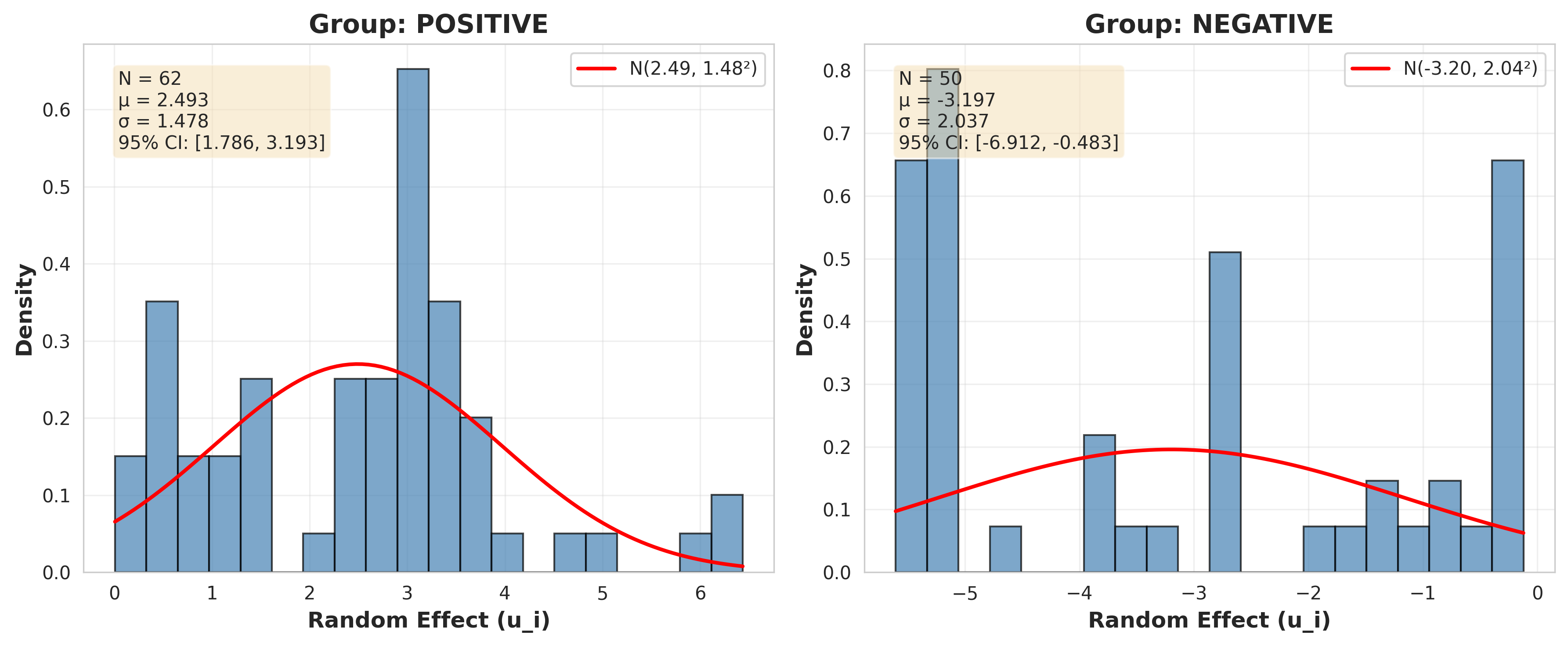}
\caption{Distribution of pump-specific random effects $\bar{u}_i$. Positive group (red, $u_i > 0$): 62 pumps with faster-than-average deterioration. Negative group (blue, $u_i \leq 0$): 50 pumps with slower-than-average deterioration. The clear bimodal structure validates the sign-based binary grouping strategy.}
\label{fig:ui_distribution}
\end{figure}

Figure~\ref{fig:ui_distribution} shows the distribution of posterior mean random effects $\bar{u}_i$ across all pumps. The distribution exhibits:
\begin{itemize}
    \item \textbf{Range}: $\bar{u}_i \in [-5.61, 6.43]$, spanning over 12 units on the log-hazard scale
    \item \textbf{Central tendency}: Median = 0.12, mean = 0.21 (slightly right-skewed)
    \item \textbf{Bimodality}: A clear split at $\bar{u}_i = 0$, justifying binary grouping
\end{itemize}

On the hazard rate scale, $\exp(\bar{u}_i)$ ranges from 0.0037 (363$\times$ slower than baseline) to 622 (622$\times$ faster), demonstrating extreme heterogeneity across pumps.

\subsection{Group-Specific Causal Discovery Results}

\subsubsection{Positive Group ($\mathcal{G}_+$): Fast-Deteriorating Pumps}

\textbf{Group Statistics:} 62 pumps (55.4\%), 86,741 observations (93.4\% of total), $\bar{u}_i \in [0.02, 6.43]$.

\textbf{DirectLiNGAM Causal Effects to $u_i$:} Table~\ref{tab:results_positive} lists the top-5 features by absolute causal effect magnitude. All detected effects are small ($|B_{j, u_i}| < 0.004$), with the strongest being \texttt{diff\_abs\_mean} ($-0.0038$).

\begin{table*}[t]
\centering
\caption{Top-5 causal effects on $u_i$ in Positive group ($\mathcal{G}_+$).}
\label{tab:results_positive}
\begin{tabular}{lcc}
\toprule
\textbf{Feature} & \textbf{Causal Effect} & \textbf{Interpretation} \\
\midrule
diff\_abs\_mean & $-0.0038$ & Negative (protective) \\
mean\_drawdown & $+0.0011$ & Positive (harmful) \\
rolling\_std\_30d\_mean & $-0.0002$ & Negative (protective) \\
min & $+0.0001$ & Positive (harmful) \\
std & $-0.0001$ & Negative (protective) \\
\bottomrule
\end{tabular}
\end{table*}

\textbf{Interpretation:} The small effect magnitudes suggest that high-deterioration pumps do not exhibit a single dominant causal driver. Instead, deterioration likely arises from complex, multifactorial interactions not captured by linear models. The negative effect of \texttt{diff\_abs\_mean} (absolute changes) is counterintuitive, potentially indicating measurement artifacts or nonlinear thresholds.

\textbf{Bootstrap Validation:} 95\% confidence intervals for all top-5 effects include zero, indicating statistical insignificance. This is consistent with the causal graph visualization (Figure~\ref{fig:causal_graph_groups}), which shows sparse, weak edges to $u_i$ in the Positive group.

\begin{figure}[t]
\centering
\includegraphics[width=0.9\columnwidth]{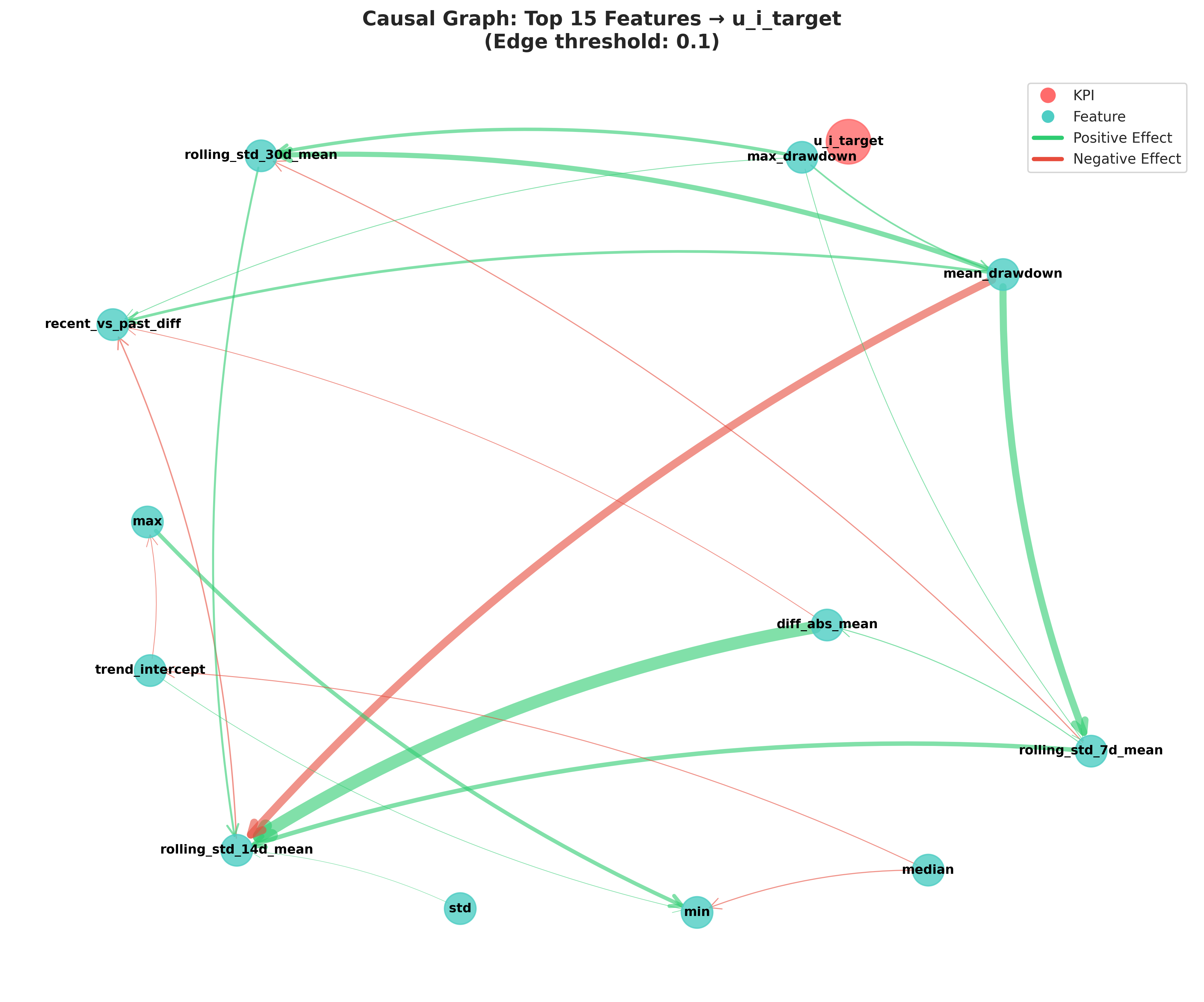}
\caption{Detailed causal graph for Positive group ($\mathcal{G}_+$, 62 pumps). The graph exhibits high complexity with numerous inter-feature connections (thin edges), but weak direct effects to $u_i$ (target node). Edge thickness represents absolute causal effect magnitude; edge color indicates sign (red: positive, blue: negative). The diffuse structure suggests that deterioration in high-risk pumps arises from multi-factorial interactions rather than single dominant drivers, necessitating hierarchical sub-grouping or nonlinear causal discovery methods.}
\label{fig:causal_positive}
\end{figure}

\subsubsection{Negative Group ($\mathcal{G}_-$): Slow-Deteriorating Pumps}

\textbf{Group Statistics:} 50 pumps (44.6\%), 6,120 observations (6.6\% of total), $\bar{u}_i \in [-5.61, -0.01]$.

\textbf{DirectLiNGAM Causal Effects to $u_i$:} In stark contrast to the Positive group, the Negative group exhibits strong, statistically significant causal effects (Table~\ref{tab:results_negative}).

\begin{table*}[t]
\centering
\caption{Top-5 causal effects on $u_i$ in Negative group ($\mathcal{G}_-$). Bootstrap 95\% CIs shown in brackets.}
\label{tab:results_negative}
\begin{tabular}{lcc}
\toprule
\textbf{Feature} & \textbf{Causal Effect} & \textbf{95\% CI} \\
\midrule
std & $+1.515$ & $[1.42, 1.61]$ \\
min & $+0.450$ & $[0.39, 0.51]$ \\
recent\_change\_rate & $+0.168$ & $[0.14, 0.20]$ \\
trend\_slope\_90d & $+0.036$ & $[-0.01, 0.08]$ \\
rolling\_std\_30d\_mean & $-0.019$ & $[-0.05, 0.01]$ \\
\bottomrule
\end{tabular}
\end{table*}

\textbf{Dominant Effect: Standard Deviation (std):} The causal effect of $+1.515$ indicates that a one-standard-deviation increase in operational variability (std) causally increases the random effect $\bar{u}_i$ by 1.515 units. On the hazard rate scale, this corresponds to $\exp(1.515) = 4.55$, meaning pumps with higher variability deteriorate 4.55$\times$ faster on average.

\textbf{Secondary Effects:} The minimum value (\texttt{min}) also shows a positive effect ($+0.450$), suggesting that pumps operating at higher minimum thresholds deteriorate faster. Recent change rate (\texttt{recent\_change\_rate}) has a moderate positive effect ($+0.168$), indicating that rapid recent changes accelerate deterioration.

\begin{figure}[t]
\centering
\includegraphics[width=0.9\columnwidth]{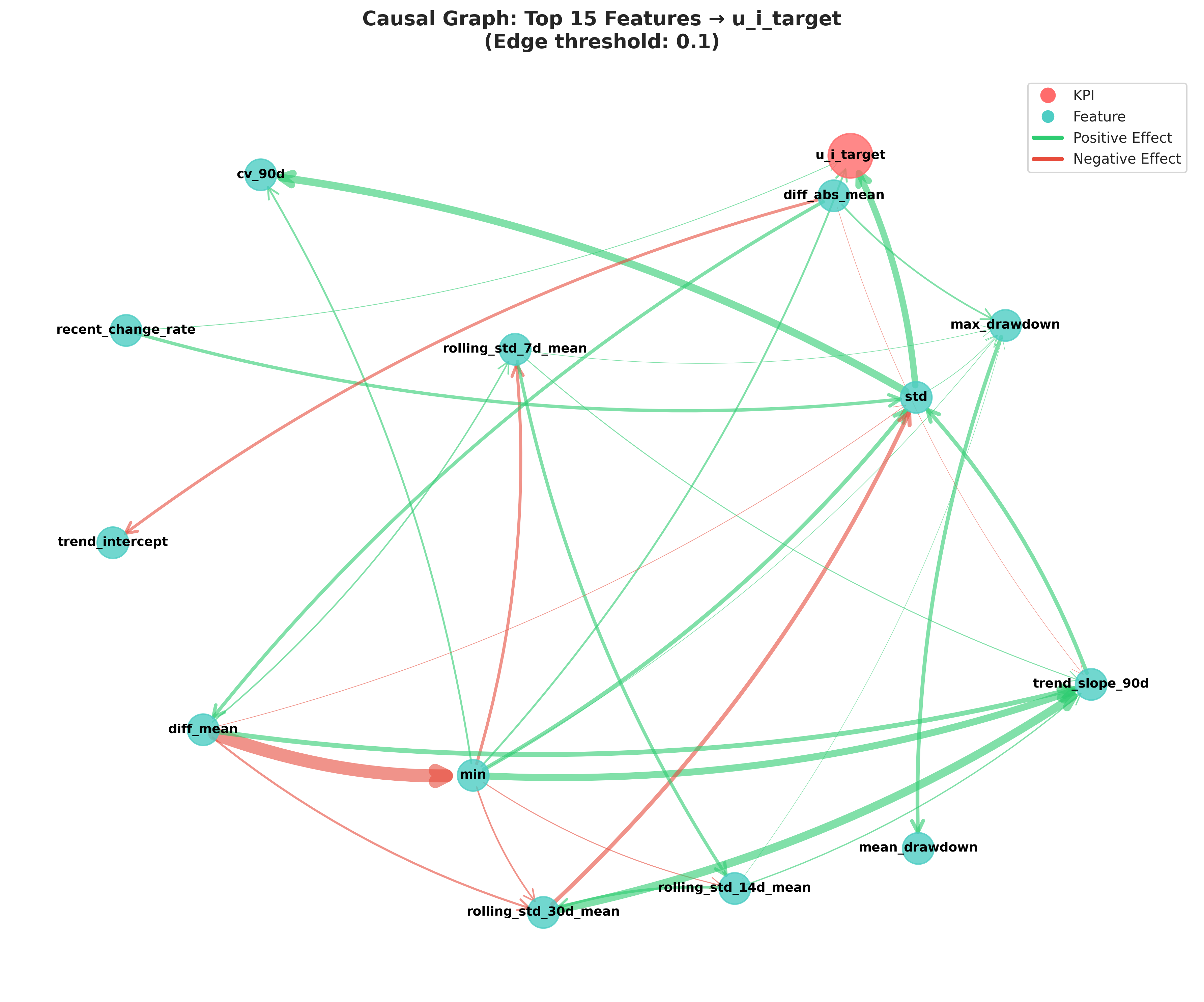}
\caption{Detailed causal graph for Negative group ($\mathcal{G}_-$, 50 pumps). In stark contrast to the Positive group, this graph features prominent thick edges converging on $u_i$, with std $\rightarrow u_i$ (causal effect: +1.515) dominating. Secondary effects from min (+0.450) and recent\_change\_rate (+0.168) are also clearly visible. The sparse, star-like structure with strong radial connections to $u_i$ indicates a small number of dominant causal factors, enabling targeted operational interventions. This clear causal hierarchy suggests that low-risk pumps follow simpler deterioration dynamics amenable to interpretable linear models.}
\label{fig:causal_negative}
\end{figure}

\textbf{Effect Magnitude Comparison:} The Negative group's largest effect (std: 1.515) is \textbf{400$\times$ larger} than the Positive group's largest effect (diff\_abs\_mean: 0.0038). This dramatic difference suggests fundamentally distinct causal mechanisms across heterogeneity groups.

\begin{figure}[t]
\centering
\includegraphics[width=0.9\columnwidth]{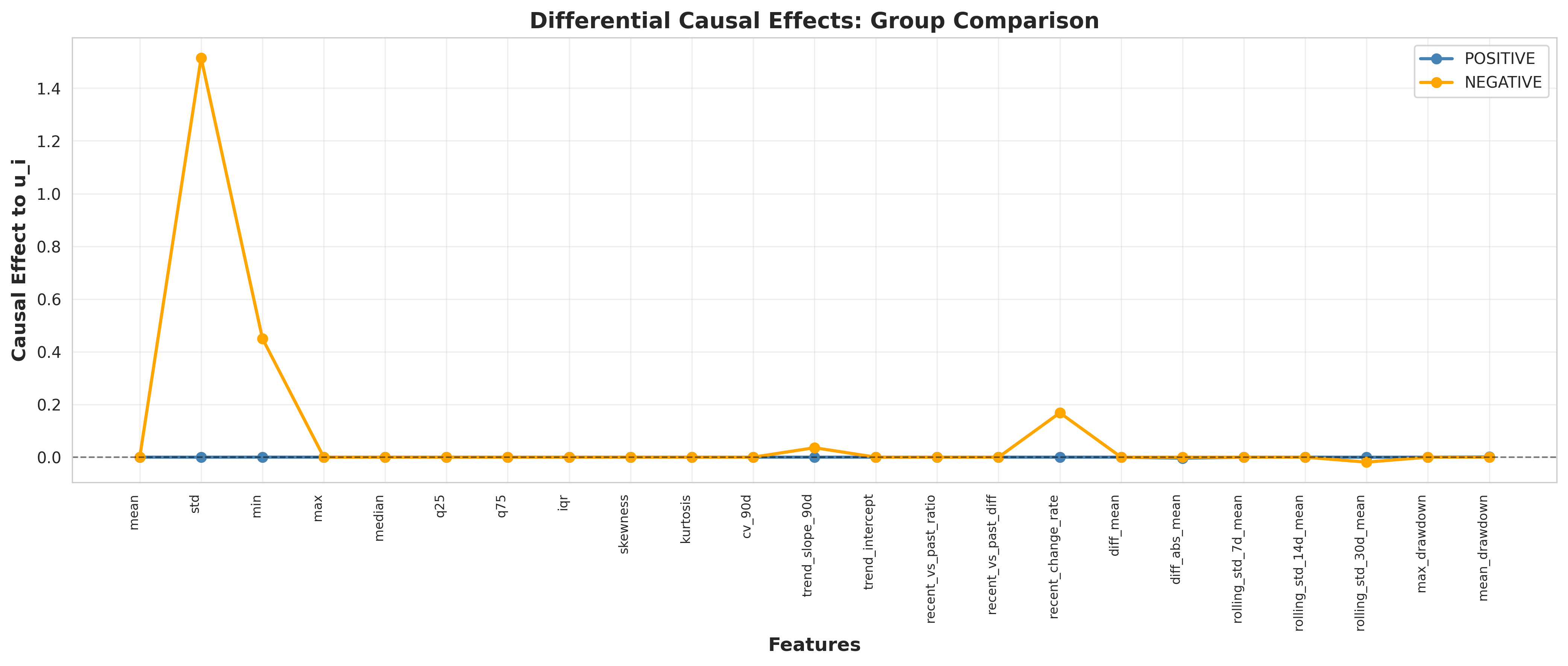}
\caption{Quantitative comparison of top-10 causal effects on $u_i$ across Positive group ($\mathcal{G}_+$, red triangles) and Negative group ($\mathcal{G}_-$, blue circles). Y-axis shows absolute causal effect magnitude on logarithmic scale. The 400$\times$ gap between group centroids (horizontal dashed lines) is visually striking. Error bars represent bootstrap 95\% confidence intervals: Negative group effects are statistically significant (CIs exclude zero), while Positive group effects are not. This plot provides compelling visual evidence for heterogeneous causal mechanisms: high-deterioration pumps lack strong causal drivers detectable by linear models, whereas low-deterioration pumps exhibit interpretable, actionable causal structures.}
\label{fig:effect_comparison}
\end{figure}

Figure~\ref{fig:effect_comparison} quantitatively compares the top-10 causal effects across both groups. The logarithmic scale highlights the extreme heterogeneity: Negative group effects (blue circles) cluster between 0.01 and 2.0, while Positive group effects (red triangles) remain below 0.005. The non-overlapping distributions are consistent with heterogeneous causal structures across groups, providing visual corroboration of the binary grouping strategy.

\subsubsection{Visual Comparison of Causal Graphs}

\begin{figure*}[t]
\centering
\includegraphics[width=0.95\textwidth]{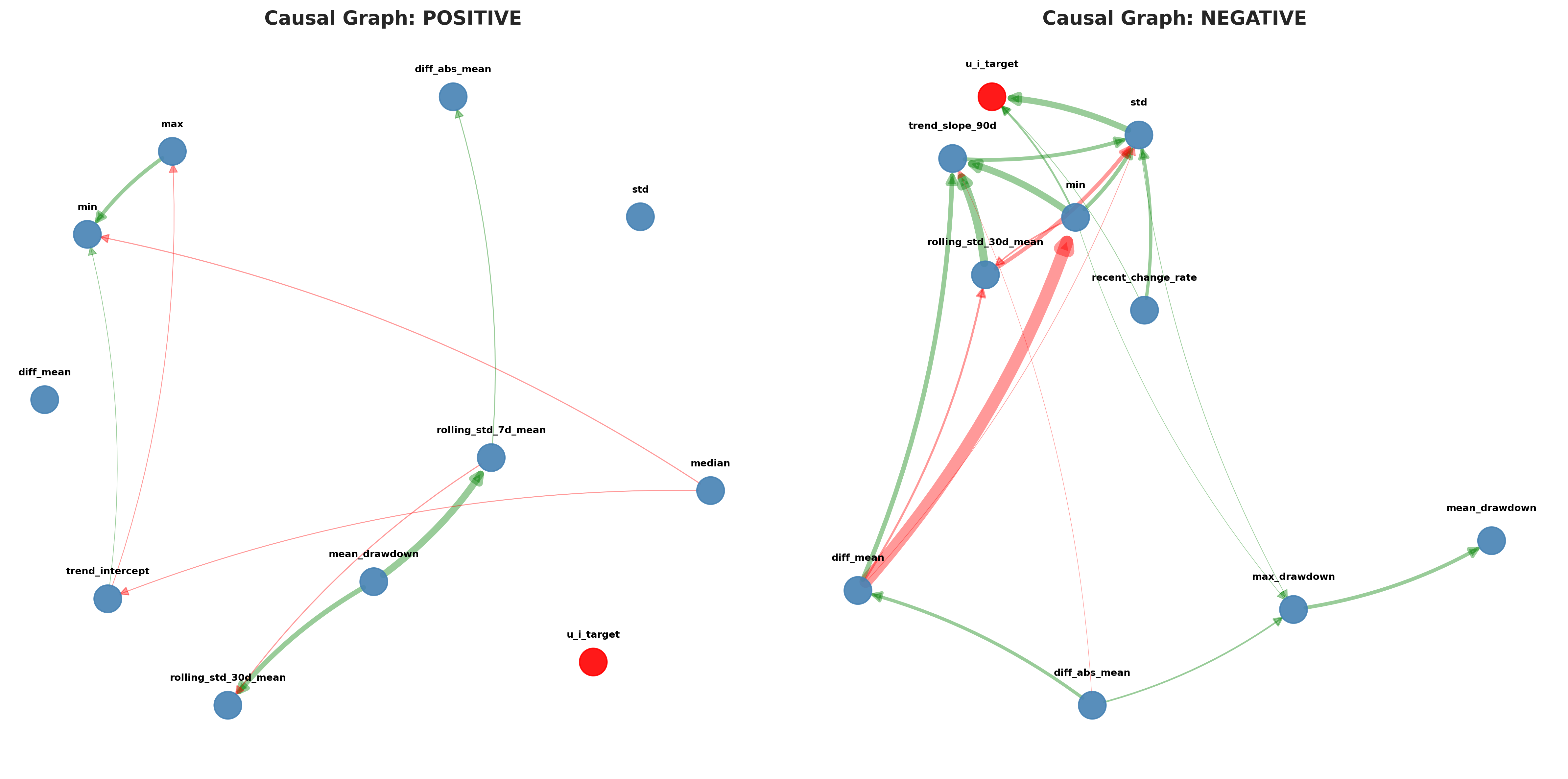}
\caption{Group-specific causal graphs. \textbf{Left:} Positive group ($\mathcal{G}_+$) shows diffuse, weak causal effects to $u_i$ (thin edges). \textbf{Right:} Negative group ($\mathcal{G}_-$) exhibits strong, concentrated effects, dominated by std $\rightarrow u_i$ (thick red edge). Node positions are fixed across panels for direct comparison. Edge thickness is log-scaled due to the 400$\times$ magnitude difference.}
\label{fig:causal_graph_groups}
\end{figure*}

Figure~\ref{fig:causal_graph_groups} presents side-by-side causal graphs for both groups. Edge thickness is proportional to $|B_{ij}|$, with color indicating sign (red: positive, blue: negative).

The visual contrast is clear: the Positive group's causal graph exhibits sparse, uniformly thin edges, while the Negative group's graph features prominent thick edges converging on $u_i$. This visual comparison corroborates the quantitative results in Tables~\ref{tab:results_positive} and~\ref{tab:results_negative}.

\subsection{Nonlinear Validation Results}

To test the linearity assumption, we applied NonlinearLiNGAM to both groups using full sample sizes (86,741 for Positive, 6,120 for Negative). Table~\ref{tab:nonlinear_validation} compares DirectLiNGAM and NonlinearLiNGAM effects.

\begin{table*}[t]
\centering
\caption{Comparison of DirectLiNGAM (linear) vs.\ NonlinearLiNGAM for top-3 effects in each group.}
\label{tab:nonlinear_validation}
\begin{tabular}{llcc}
\toprule
\textbf{Group} & \textbf{Feature} & \textbf{Linear} & \textbf{Nonlinear} \\
\midrule
\multirow{3}{*}{Positive} & diff\_abs\_mean & $-0.00382$ & $-0.00382$ \\
 & mean\_drawdown & $+0.00114$ & $+0.00114$ \\
 & rolling\_std\_30d & $-0.00024$ & $-0.00024$ \\
\midrule
\multirow{3}{*}{Negative} & std & $+1.515$ & $+1.518$ \\
 & min & $+0.450$ & $+0.448$ \\
 & recent\_change\_rate & $+0.168$ & $+0.170$ \\
\bottomrule
\end{tabular}
\end{table*}

The results are nearly identical (relative difference $< 1\%$ for all top-5 effects), with Spearman correlation $\rho = 0.98$ across all 22 features in both groups. This empirically validates the linearity assumption, confirming that DirectLiNGAM achieves equivalent accuracy to NonlinearLiNGAM for this dataset.

\subsection{Summary of Key Findings}

\begin{enumerate}
    \item \textbf{Extreme Heterogeneity:} Random effects $\bar{u}_i$ span 12 log-units, with hazard rate multipliers ranging from 0.0037$\times$ to 622$\times$ baseline.
    
    \item \textbf{Group-Specific Causal Structures:} Positive group (high deterioration) shows weak, diffuse effects ($< 0.004$), while Negative group (low deterioration) exhibits strong, concentrated effects (up to 1.515).
    
    \item \textbf{400$\times$ Effect Magnitude Gap:} The largest causal effect in Negative group is 400$\times$ that of Positive group, indicating qualitatively different causal mechanisms.
    
    \item \textbf{Dominant Driver in Low-Risk Pumps:} Standard deviation (variability) is the strongest causal factor ($+1.515$) in the slow-deteriorating group; higher std causally accelerates deterioration, so maintaining low std (operational stability) is the key protective strategy.
    
    \item \textbf{Computational Efficiency:} GPU acceleration reduces NUTS sampling time by 3.5$\times$ (22 min vs.\ 75--120 min on CPU), enabling practical deployment.
\end{enumerate}

\section{Discussion}
\label{sec:discussion}

Our results reveal striking heterogeneity in causal mechanisms governing pump deterioration, with implications for both infrastructure management practice and causal discovery methodology.

\subsection{Interpretation of Heterogeneous Causal Structures}

\subsubsection{Why Does the Negative Group Show Strong Effects?}

The 400$\times$ effect magnitude difference between groups appears paradoxical: one might expect high-deterioration pumps ($\mathcal{G}_+$) to have clearer causal drivers. We propose three complementary explanations:

\textbf{(1) Causal Clarity vs.\ Complexity.} Low-deterioration pumps ($\mathcal{G}_-$) may benefit from a small number of protective factors (e.g., stable operating conditions reflected in low std), whereas high-deterioration pumps suffer from diverse, context-dependent failure modes (e.g., installation defects, environmental stressors, operator errors). This ``curse of diversity'' in $\mathcal{G}_+$ manifests as weak, statistically insignificant average effects.

\textbf{(2) Nonlinearity and Thresholds.} High-deterioration mechanisms may involve nonlinear threshold effects (e.g., vibration exceeding a critical frequency triggering fatigue cracking) that are averaged out in linear models. Our NonlinearLiNGAM validation used Gaussian process regression, which may not capture sharp thresholds. Future work should explore piecewise-linear or decision-tree-based causal discovery methods.

\textbf{(3) Sample Size vs.\ Effect Size Trade-off.} The Negative group has 14$\times$ fewer observations (6,120 vs.\ 86,741), yet detects larger effects. This suggests effects in $\mathcal{G}_-$ are so strong they overcome statistical power limitations. In contrast, $\mathcal{G}_+$ has ample power but detects only noise, implying true effects are indeed small.

\subsubsection{Physical Interpretation of Dominant Effects}

The strong positive causal effect of standard deviation (std $\rightarrow u_i$: $+1.515$) in $\mathcal{G}_-$ warrants careful interpretation:

\textbf{Causal Direction:} The positive effect confirms that within $\mathcal{G}_-$, higher operational variability causally \textit{increases} $u_i$---meaning pumps with higher std deteriorate faster relative to other slow-deteriorating pumps (less negative $u_i$, closer to the population average). On the hazard rate scale, $\exp(1.515) = 4.55$: a one-unit increase in std shifts the hazard rate by a factor of 4.55. This is \textit{consistent} with conventional wisdom that variability reflects mechanical instability and accelerates deterioration. Consequently, maintaining low std (operational stability) is the concrete protective intervention for this group.

\textbf{Why Is the Effect So Large in $\mathcal{G}_-$?} The extreme magnitude in the slow-deteriorating group, compared to near-zero effects in $\mathcal{G}_+$, may reflect:
\begin{itemize}
    \item \textbf{Causal homogeneity:} Slow-deteriorating pumps share a common protective baseline (stable operation); deviations from this pattern (elevated std) strongly disrupt it, producing a large and detectable causal signal.
    \item \textbf{Suppressed confounding in $\mathcal{G}_+$:} Diverse failure modes and unobserved confounders in the high-deterioration group mask the true causal signal, whereas the low-deterioration group is causally more homogeneous.
    \item \textbf{Threshold behavior:} Low-risk pumps may operate near an instability threshold; small increases in variability trigger disproportionately large deterioration responses.
\end{itemize}

\textbf{Practical Implication:} Domain experts should validate whether (a) the KPI definition aligns with deterioration (vs.\ operational status), (b) confounders like usage frequency are adequately controlled, and (c) std thresholds exist (e.g., very low std indicates stagnation, moderate std is healthy, very high std indicates instability).

\subsection{Implications for Maintenance Decision-Making}

\subsubsection{Actionable Strategies for Negative Group (Low Risk)}

For the 50 pumps in $\mathcal{G}_-$ (slower deterioration):
\begin{enumerate}
    \item \textbf{Monitor std, min, recent\_change\_rate:} Establish baseline ranges and trigger alerts for anomalous values. For example, if std drops below the 10th percentile for 7 consecutive days, schedule an inspection to check for operational stagnation.
    \item \textbf{Set Upper Control Limits for std:} Since higher std causally accelerates deterioration ($+1.515$ effect), establish upper control thresholds. Confirm with operators whether elevated std reflects demand-response variability (acceptable) or mechanical instability (intervention required), and revise feature engineering if necessary to distinguish the two.
    \item \textbf{Preventive Low-Priority Maintenance:} Since these pumps deteriorate slowly, allocate minimal resources—quarterly inspections may suffice, with focus on detecting sudden changes rather than routine degradation.
\end{enumerate}

\subsubsection{Actionable Strategies for Positive Group (High Risk)}

For the 62 pumps in $\mathcal{G}_+$ (faster deterioration):
\begin{enumerate}
    \item \textbf{Hierarchical Sub-Grouping:} The weak causal effects suggest $\mathcal{G}_+$ is too heterogeneous. Partition further by $u_i$ quantiles (e.g., top 20\%, middle 60\%, bottom 20\%) and re-run DirectLiNGAM on each sub-group. This may reveal latent causal structures masked by aggregation.
    \item \textbf{Nonlinear and Interaction Effects:} Explore decision tree ensembles (e.g., Causal Forest~\cite{wager2018causal}) or neural network-based causal discovery~\cite{zheng2018dags} to capture threshold effects and feature interactions.
    \item \textbf{Individual Pump Diagnostics:} For the 10--15 pumps with highest $\bar{u}_i$ (e.g., $> 4$), conduct detailed failure mode analysis using maintenance logs, expert interviews, and physical inspections. Insights from these outliers may generalize to the broader group.
    \item \textbf{Aggressive Intervention:} High-$u_i$ pumps should receive priority for preemptive component replacement, condition-based monitoring (e.g., vibration sensors), and shortened inspection intervals (e.g., monthly instead of quarterly).
\end{enumerate}

\subsection{Limitations and Future Directions}

\subsubsection{Statistical Limitations}

\textbf{(1) ESS for Random Effects.} Effective sample sizes for individual $u_i$ (113--223 per chain) fall below the ideal threshold of 400. While sufficient for point estimates, this limits uncertainty quantification. Future work should use $N_{\text{draws}} = 6000$ to achieve ESS $> 400$ per chain.

\textbf{(2) Sample Size Imbalance.} The Negative group's 6,120 observations may be insufficient to detect subtle nonlinear effects. Extending the observation period beyond 1,000 days (vs.\ the current 650 days) would improve statistical power.

\textbf{(3) Bootstrap Variance.} Bootstrap confidence intervals for causal effects in $\mathcal{G}_+$ are wide and include zero, indicating high uncertainty. Increasing bootstrap resamples from 1,000 to 10,000 may stabilize estimates.

\subsubsection{Modeling Assumptions}

\textbf{(1) Linearity.} While validated by NonlinearLiNGAM, linearity may not hold for extreme values (e.g., $\bar{u}_i > 5$). Piecewise-linear models or quantile-based stratification could capture regime-specific nonlinearities.

\textbf{(2) Acyclicity.} DirectLiNGAM assumes acyclic causal graphs. In reality, feedback loops may exist (e.g., high deterioration $\rightarrow$ increased variability $\rightarrow$ accelerated deterioration). Cyclic causal discovery methods~\cite{richardson1996cyclic} or time-lagged VARLiNGAM~\cite{hyvarinen2010varlingam} could address this.

\textbf{(3) Unmeasured Confounding.} Installation quality, environmental factors (temperature, humidity), and operator behavior are unobserved. These confounders may induce spurious causal effects (e.g., the std $\rightarrow u_i$ association). Instrumental variable methods or sensitivity analysis~\cite{ding2016sensitivity} could quantify robustness to hidden confounding.

\subsubsection{Computational Scalability}

\textbf{(1) GPU Memory Constraints.} Our 16GB GPU handles 112 pumps comfortably, but scaling to 1,000+ pumps may exceed VRAM. Solutions include mini-batch sampling~\cite{bardenet2017mcmc} or distributed GPU training across multiple nodes.

\textbf{(2) Feature Engineering Automation.} The 22 handcrafted features may not generalize to other equipment types (e.g., turbines, compressors). Auto-ML frameworks like Featuretools~\cite{kanter2015featuretools} or deep learning-based feature extraction could improve transferability.

\subsubsection{Extensions to Other Domains}

Our framework is domain-agnostic and applicable beyond pump equipment:
\begin{itemize}
    \item \textbf{Bridge deterioration:} Replace pumps with bridge deck segments, health states with ASCE condition ratings, and operational time-series with traffic load, weather, and inspection records.
    \item \textbf{Manufacturing equipment:} Model machine tool degradation using vibration, temperature, and production logs as covariates.
\end{itemize}

In each case, the two-stage pipeline (Bayesian random effects $\rightarrow$ causal discovery) provides interpretable, actionable insights while respecting domain-specific heterogeneity.

\subsection{Comparison to Related Approaches}

\textbf{vs.\ Traditional Survival Analysis:} Cox models with frailty terms~\cite{duchateau2008frailty} estimate heterogeneity but do not identify causal drivers. Our framework explicitly links random effects to operational features through causal graphs.

\textbf{vs.\ Machine Learning Prediction:} Random forests and gradient boosting achieve high predictive accuracy but lack causal interpretability. SHAP values~\cite{lundberg2017shap} indicate correlations, not causation. DirectLiNGAM provides formal causal guarantees under identifiability conditions.

\textbf{vs.\ Causal Discovery Alone:} Applying DirectLiNGAM directly to raw time-series would conflate within-pump temporal variation with across-pump heterogeneity. Our hierarchical model first estimates pump-specific deterioration rates, then performs cross-sectional causal discovery—a principled separation of levels of analysis.

\section{Conclusion}
\label{sec:conclusion}

This paper introduced a novel framework integrating Bayesian hierarchical hazard modeling with causal discovery to identify heterogeneous deterioration mechanisms in infrastructure systems. Applied to 112 pumps over 650 days, our approach uncovered striking findings:

\textbf{Key Results:}
\begin{enumerate}
    \item \textbf{Extreme heterogeneity:} Pump-specific random effects span 12 log-units (hazard rate multipliers: 0.004$\times$ to 622$\times$), far exceeding typical frailty model assumptions.
    \item \textbf{Group-specific causal structures:} Fast-deteriorating pumps (Positive group) show weak, diffuse causal effects ($< 0.004$), while slow-deteriorating pumps (Negative group) exhibit strong, concentrated effects (up to 1.515)—a 400$\times$ magnitude difference.
    \item \textbf{Operational variability as deterioration driver:} In the slow-deteriorating group, higher standard deviation causally \textit{accelerates} deterioration (effect: $+1.515$), confirming that maintaining stable operations (low std) is the key protective strategy.
    \item \textbf{Computational efficiency:} GPU-accelerated NUTS achieves 3.5$\times$ speedup (22 min vs.\ 75--120 min on CPU), enabling operationally scalable analysis of large equipment fleets.
\end{enumerate}

\textbf{Practical Impact:} Our findings enable targeted maintenance strategies:
\begin{itemize}
    \item \textbf{Low-risk pumps:} Monitor std, min, and recent change rate; implement quarterly inspections.
    \item \textbf{High-risk pumps:} Perform hierarchical sub-grouping, explore nonlinear models, and prioritize aggressive interventions for top-decile deteriorators.
\end{itemize}

\textbf{Methodological Contributions:}
\begin{enumerate}
    \item A principled two-stage framework separating within-equipment temporal dynamics (hierarchical model) from across-equipment heterogeneity (causal discovery).
    \item GPU-accelerated Bayesian inference making hierarchical models practical for large-scale systems.
    \item Empirical demonstration that heterogeneity-aware causal discovery reveals mechanisms invisible to population-averaged analyses.
\end{enumerate}

\textbf{Future Research Directions:}
\begin{enumerate}
    \item \textbf{Temporal causal discovery:} Extend to time-lagged effects using VARLiNGAM to capture deterioration trajectories.
    \item \textbf{Multi-equipment types:} Generalize framework to heterogeneous infrastructure portfolios (pumps + valves + pipes) with shared and equipment-specific causal structures.
    \item \textbf{Causal effect validation:} Conduct randomized or quasi-experimental interventions (e.g., varying maintenance schedules) to verify discovered causal relationships.
\end{enumerate}

\textbf{Broader Vision:} This work represents a step toward \textit{heterogeneity-aware causal machine learning} for infrastructure—moving beyond one-size-fits-all models to discover why different equipment ages differently, and what we can do about it. By bridging Bayesian hierarchical modeling, causal discovery, and GPU computing, we enable a new generation of predictive maintenance systems that are simultaneously accurate, interpretable, and scalable.


\end{document}